\documentclass{appolb}
\usepackage{epsfig}

\newcommand{\lsim}{\raisebox{-0.13cm}{~\shortstack{$<$ \\[-0.07cm] $\sim$}}~}
\newcommand{\gsim}{\raisebox{-0.13cm}{~\shortstack{$>$ \\[-0.07cm] $\sim$}}~}
\newcommand{\bea}{\begin{eqnarray}}
\newcommand{\eea}{\end{eqnarray}}

\begin{document}
\title{NEUTRINO MIXING AND LEPTON FLAVOR VIOLATION IN SUSY-GUT MODELS%
\thanks{Presented by J.I.I. at the XXVII International Conference of 
Theoretical Physics, Ustro\'n, Poland, September 15-21, 2003.}%
}
\author{Jos\'e I. Illana and Manuel Masip
\address{Centro Andaluz de F\'\i sica de Part\'\i culas Elementales (CAFPE)\\ 
Departamento de F\'\i sica Te\'orica y del Cosmos \\
Universidad de Granada, E-18071 Granada, Spain}
}
\maketitle
\vspace{-6.5cm}
\begin{flushright}
CAFPE-27/03\\
UG-FT-157/03
\end{flushright}
\vspace{4.5cm}
\begin{abstract}
In supersymmetric (SUSY) models the misalignment between fermion 
and sfermion families introduces unsuppressed flavor-changing 
processes. Even if the mass parameters are chosen to give no flavor 
violation, family dependent radiative corrections make this adjustment 
not stable.
In particular, due to the observed large neutrino mixings and 
potentially large neutrino Yukawa couplings, sizable lepton 
flavor violation (LFV) is expected. 
After introducing the basic concepts, the framework and the 
main assumptions, we report on a recent study of rare 
leptonic decays in a class of SUSY-GUT models with three 
quasi-degenerate neutrinos. We show that LFV effects are
likely visible in forthcoming experiments.
\end{abstract}
  
\section{Lepton Flavor Violation}

In the standard model (SM) 
neutrinos are massless and hence 
individual lepton flavor is conserved. This is in contrast with
the quark sector, where the 
Cabibbo-Kobayashi-Maskawa (CKM) mixing allows for 
(one-loop, GIM suppressed) flavour changing neutral currents.
In particular, the process $b\to s\gamma$ is predicted at 
a rate in agreement with experiment.

The observation of neutrino oscillations has opened for the first
time the gate to physics beyond the SM. It requires the introduction
of right-handed neutrino isosinglets and 
Dirac mass terms analogous to the ones in the quark sector. 
Majorana mass terms for the singlets are also possible (although
not indispensable), and they could be used to explain the lightness of 
the neutrino masses via the seesaw 
mechanism. In either case, LFV processes like $\mu\to e\gamma$, 
$Z\to\mu e$, $\mu\to3e$ are possible at a tiny rate in this so-called
$\nu$SM. Thus
\bea
{\rm\nu SM}:\ \mbox{BR}(\ell_i\to\ell_j\gamma)\sim\frac{\alpha^3}{G^2_F}
\frac{|M^*_{\nu_{ik}}M_{\nu_{kj}}|^2}{M^8_W}\lsim 10^{-45}\ ,
\eea
where the neutrino mass matrix
${\bf M_\nu}={\bf U}^*\mbox{diag}(m_1,m_2,m_3){\bf U}^\dagger$ 
contains off-diagonal entries 
due to the misalignment between the charged-lepton and neutrino 
mass matrices encoded in the Maki-Nakagawa-Sakata (MNS) mixing 
matrix ${\bf U}$. Present experimental constrains 
$M_{\nu_{ij}}\lsim1$~eV are used.

SUSY extensions of the SM introduce new sources of flavor 
non-conser\-vation: the misalignment of lepton and slepton mass matrices
causes unsuppressed LFV. The dominant contribution in the minimal 
supersymmetric standard model (MSSM) is
\bea
{\rm MSSM}:\ \mbox{BR}(\ell_i\to\ell_j\gamma)\sim\frac{\alpha^3}{G^2_F}
\frac{|m^2_{\tilde L_{ij}}|^2}{m^8_S}\tan^2\beta\lsim
(\delta^{ij}_{LL})^2\tan^2\beta\times10^{-5}\ ,
\eea
where the off-diagonal (in the basis of charged-lepton mass eigenstates)
slepton mass insertions 
$m^2_{\tilde L_{ij}}\sim m^2_S\delta_{Lij}$ are normalized to a typical SUSY
mass scale $m_S\gsim10^2$~GeV. Even for low $\tan\beta$, the experimental
limits require a very fine alignment, in particular for the first two
generations: BR$(\mu\to e\gamma)\lsim 10^{-11}\Rightarrow\delta^{12}_{LL}\lsim
10^{-3}$. This is generally formulated as the SUSY {\em flavor problem}: how 
to explain such an alignment when fermion and sfermion masses have a
different origin; the former come from (SUSY-preserving) Yukawa 
interactions ($Y_{ij}$) and the latter from a SUSY-breaking mechanism 
($m^2_{ij}$).

The most economical solution is that ${\bf m^2\propto 1}$. This can be 
the case in SUGRA scenarios, 
where SUSY is broken in a hidden sector only connected
via gravitational interactions with the SM. Gravitational interactions
could generate
identical masses for the three sfermion masses at the Planck scale. But the
renormalization-group (RG) corrections from the Planck down to the electroweak
scale would then introduce off-diagonal entries due to the different
Yukawa interactions of the three slepton families. 
The impact of this effect on LFV decays is the subject of 
our analysis \cite{Illana:2003pj}, with special emphasis on the off-diagonal 
slepton mass terms generated from the Planck to the GUT scale, usually 
neglected in the literature.

\section{Seesaw and neutrino Yukawa couplings}

The seesaw mechanism 
requires the addition of one gauge singlet ($N_i$) to
each SM doublet ($\nu_i$), and needs that Majorana
masses (${\bf M}$) of the singlets 
are much larger than Dirac mass terms (${\bf M_D}$). 
As a consequence one gets, apart from heavy neutrinos with masses of 
order ${\bf M}$, light neutrinos with masses 
${\bf M_\nu}={\bf M_D^{\rm T}}{\bf M}^{-1}{\bf M_D}$, where 
${\bf M_D}=\langle H_2\rangle{\bf Y_\nu}$ is obtained via Yukawa
interactions after the electroweak symmetry breaking when the hypercharge 
$+1/2$ Higgs field ($H_2$ in the MSSM) gets a {\em vev} 
$\langle H_2\rangle=v_2=v\sin\beta$. Notice that a mass of the order of
1~eV can be generated by a coupling $Y_\nu\sim1$ with $M\sim10^{14}$~GeV
or by $Y_\nu\sim10^{-2}$ with $M\sim10^{10}$~GeV.

Neutrino oscillations give us a partial information 
\cite{Gonzalez-Garcia:2003qf} on neutrino
masses (the squared mass differences $\Delta m^2_{21}$ and $\Delta m^2_{32}$) 
and the neutrino mixing 
matrix ${\bf U}$ ($\theta_{12},\theta_{23},\theta_{13}$).
In the basis where the charged-lepton Yukawa matrix ${\bf Y_e}$ is 
diagonal, ${\bf U}$ relates 
the weak eigenstates $\nu_\alpha$ ($\alpha=e,\mu,\tau$)
and the mass eigenstates $\nu_i$ ($i=1,2,3$): 
$\nu_\alpha=U_{\alpha i}\nu_i$.

The neutrino Yukawa matrix can be expressed \cite{Casas:2001sr} in a basis 
where the Majorana mass matrix is diagonal (${\bf D_M}$) in terms of the
Majorana masses, the diagonal light neutrino mass matrix (${\bf D_m}$) 
and the neutrino mixings up to an unknown complex orthogonal matrix ${\bf R}$,
\bea
v_2{\bf Y_\nu}=\sqrt{\bf D_M}{\bf R}\sqrt{\bf D_m}{\bf U}^\dagger\ .
\eea
Therefore, it depends on six masses, six angles and six phases. Only
a few of these parameters are accessible to current experiments. To simplify
the analysis we adopt the following {\em conservative} assumptions (they
will lead to a minimal LFV): (1)~We concentrate on the case of 
quasi-degenerate (QD) neutrinos (${\bf D_m}\approx m_\nu {\bf 1}$), that
{\em naturally} implies ${\bf D_M} = M {\bf 1}$. (2)~We assume that ${\bf R}$
is real, that implies 
that $v^2_2{\bf Y^\dagger_\nu Y_\nu}=M{\bf UD_mU^\dagger}$
(relevant for LFV) is independent of ${\bf R}$. It must be noticed that
a complex ${\bf R}$ is necessary for leptogenesis and enhances LFV by
several orders of magnitude \cite{Pascoli:2003rq}. (3)~Finally we assume
that CP is conserved and the light neutrinos have all the same CP parities
(${\bf U}$ is real and the eigenvalues of ${\bf M_\nu}$ have all the same sign).
For the light neutrino mass scale we take $m_\nu\approx 0.2$~eV, compatible
with $m_\nu<2.2$~eV ($^3$H $\beta-$decay), $m_\nu<0.7/3$~eV (WMAP) and
$m_\nu<(0.35-1.05)$~eV ($\beta\beta_{0\nu}$) \cite{absmass}.
 
\section{Scenarios for quasi-degenerate neutrinos}

High-energy Yukawa couplings are constrained by masses and mixings 
at the electroweak scale through the RGE. The neutrino 
Yukawa couplings do not run below the Majorana mass scale $M$, where 
they are matched to the light neutrino mass matrix ${\bf M_\nu}=
v^2_2{\bf Y^{\rm T}_\nu Y_\nu}/M$. The evolution of neutrino
mixings embodied in this effective mass matrix is controlled by charged-lepton 
Yukawa couplings. Between the electroweak and the SUSY scale ($m_S$) 
the running is negligible but above $m_S$ the down-type Yukawa couplings are
enhanced if $\tan\beta\gg1$ (in particular $Y_\tau=m_\tau/(v\cos\beta)\to 1$) 
and as a consequence mixings get magnified for QD neutrinos \cite{radmagnif}. 
Therefore we distinguish two scenarios at high-energy compatible with 
the mass differences and bimaximal mixings observed at low energy,
irrespective of the Majorana scale: 
({\it a}) Bimaximal mixings for low $\tan\beta$ ($Y_\tau\ll 1$): almost no 
running; ({\it b}) CKM-like mixings for high $\tan\beta$ ($Y_\tau\sim 1$): 
from unification of quark and neutrino mixings at high scale one gets two 
large mixings $\theta_{12}$ and $\theta_{23}$. Concerning $\theta_{13}$, 
even if it is taken zero at a high scale it never vanishes at low energy, 
but becomes between $5\times10^{-5}$ and 0.1 depending on $\tan\beta$.

\section{Radiative corrections to slepton masses}

SUSY-breaking masses receive quantum
corrections from gaugino (flavor blind) and Yukawa (flavor dependent)
interactions. The sleptons (since all leptons are much lighter) get their
mass from SUSY-breaking terms:
\bea
{\bf m^2_{\tilde\ell}}\sim \pmatrix{{\bf m^2_{\tilde L}}& {\bf m^{2\dagger}_{LR}}\cr
                {\bf m^2_{LR}} & {\bf m^2_{\tilde e_R}}}\ , \quad
{\bf m^2_{\tilde\nu}}\sim {\bf m^2_{\tilde L}}\ .
\eea
Sneutrino singlets are very heavy (their masses are at least of order 
$M$) and decouple. We neglect the scalar trilinears (${\bf A_e}={\bf A_\nu}=0$) 
which in practice suppresses the left-right mixing, ${\bf m^2_{LR}}\sim 0$.
The slepton masses, taken flavor universal at tree level, 
${\bf m^2_{\tilde L}}={\bf m^2_{\tilde e_R}}=m^2_0{\bf 1}$, are changed by 
quantum corrections, developing in particular flavor-changing entries
\bea
m^2_{\tilde L_{ij}}\sim m^2_S\delta^{ij}_{LL}\ ,\quad
m^2_{\tilde e_{R\; ij}}\sim m^2_S\delta^{ij}_{RR}\ .
\eea
In the next subsections the size of the radiative corrections is 
presented in two cases: universal slepton masses at the GUT scale $M_X$,
where gauge couplings unify in the MSSM, and universality at the
Planck scale $M_P=M_{\rm Planck}/\sqrt{8\pi}$, assuming the simplest 
scenario with SU(5) gauge symmetry
between both scales. The numerical results are obtained using the
complete one-loop RG equations (see \cite{Illana:2003pj} and references
therein for more details). Analytical approximations are given below
just to understand the main effects. We define the two scenarios for
high-scale neutrino mixing described above as 
({\it a}) $\tan\beta=50$ ($Y_\tau\sim 1$) and ({\it b}) $\tan\beta=3$  
($Y_\tau\ll 1$) and consider the following cases relevant for LFV:
({\it 1}) $M=10^{14}$~GeV ($Y_\nu\sim 1$) and ({\it 2})~$M=5\times10^{11}$~GeV
($Y_\nu\ll 1$).

\subsection{Universality at the GUT scale}

In the MSSM with right-handed neutrinos the trilinear couplings of 
the superpotential in the leptonic sector are
\bea
{\cal W}=Y^{ij}_e E^c_i H_1 L_j + Y^{ij}_\nu N^c_i H_2 L_j\ ,
\eea
where $L_i$, $E^c_i$ and $N^c_i$ stand for the three families of lepton
doublets, charged singlets and neutrino singlets, respectively. $H_1$ and
$H_2$ are the MSSM Higgs-doublet superfields. 

We take diagonal charged-lepton Yukawa couplings ${\bf Y_e}$ at 
$M_X$ and include all the lepton mixing in ${\bf Y_\nu}$. 
>From universal SUSY-breaking masses at the GUT scale, 
the RG corrections introduce two relevant effects. First, the running down to
the Majorana scale $M$ generates off-diagonal terms in the slepton-doublet
mass matrix:
\bea
m^2_{\tilde L_{ij}}\approx -{3\over 8\pi^2}
m_0^2 ({\bf Y^{\dagger}_\nu Y_{\nu}})_{ij}\log{M_X \over M}\ .
\label{eq7}
\eea
The slepton-singlet mass matrix remains diagonal, 
$m^2_{\tilde e_{R\;ij}}\approx0$. 
Second, the running also generates off-diagonal terms in
${\bf Y_e}$. The low energy slepton mass matrices must be rediagonalized
accordingly. Since the rotations in the space of the three lepton
doublets $\theta^e_{ij}\approx Y^{ji}_e/Y_\tau$ ($ij=12,13,23$) are
\bea
\theta^e_{ij}\approx-\frac{1}{16\pi^2}
({\bf Y^\dagger_\nu Y_\nu})_{ij}\log\frac{M_X}{M}\ ,
\eea
and the third familiy separation (the only relevant) is
\bea
m^2_{\tilde L_{33}}-m^2_{\tilde L_{ii}}\approx-\frac{3}{8\pi^2}
m^2_0Y^2_\tau\log\frac{M_X}{M_Z}\ ,
\eea
one gets a correction $\Delta m^2_{\tilde L_{i3}}\approx
(m^2_{\tilde L_{33}}-m^2_{\tilde L_{ii}})\ \theta^e_{i3}$ to the terms 
$m^2_{\tilde L_{i3}}$ in (\ref{eq7}), which is of the order of
$Y^2_\tau/(16\pi^2)\log(M_X/M_Z)\approx 10\%$ for high $\tan\beta$.

To quantify the lepton-slepton misalignment we show the symmetric matrix
${\bf m^2_{\tilde L}}$ at the electroweak scale generated by the
given values of the scalar and gaugino masses $m_0$ and $m_{1/2}$ at $M_X$.
Only scenarios ({\it a1}) and ({\it b1}) are displayed, the 
ones where the effects are larger according to (\ref{eq7}). 
The $\delta^{ij}_{LL}$ can be read directly:
\bea
&&[m_0=m_{1/2}=300\mbox{ GeV at }M_X]\nonumber\\
\mbox{({\it a1})}\quad{\bf m^2_{\tilde L}}&=&(353 {\rm\; GeV})^2
\pmatrix{1&{-10^{-4}}&{-2\times10^{-5}}\cr
\quad*\quad&0.999&{-5\times10^{-4}}\cr
*&*&{0.793}\cr}\label{eq10}\\
\mbox{({\it b1})}\quad{\bf m^2_{\tilde L}}&=&(352 {\rm\; GeV})^2
\pmatrix{1&{-5\times10^{-5}}&{5\times10^{-5}}\cr
\quad*\quad&0.997&{-3\times10^{-3}}\cr
*&*&{0.996}\cr}\label{eq11}
\eea

\subsection{Universality at the Planck scale}

In SUSY-SU(5) each generation of quark doublets, up-quark singlets
and charged-lepton
singlets can be accommodated in the same ${\bf 10}$ irrep ($\Psi_i$)
of the group, whereas lepton doublets and down-quark singlets
would be in the ${\bf \overline 5}$ ($\Phi_i$).
We also need gauge singlets ($N^c_i$) to generate neutrino masses,
and a vectorlike ${\bf 5}+{\bf \overline 5}$ ($H_2$ and $H_1$)
to include the two standard Higgs
doublets. Other vectorlike fermions or the Higgs representations
needed to break the GUT symmetry are not essential in our
calculation. Including just the three fermion families the
trilinear terms in the superpotential read
\bea
{\cal W}_{\rm SU(5)}={1\over 4} Y_u^{ij}\; \Psi_i \Psi_j H_2 +
\sqrt{2}\ Y_{d/e}^{ij}\; \Psi_i \Phi_{j} H_{1} +
Y_{\nu}^{ij}\; N^c_i \Phi_{j} H_2\ .
\eea

At $M_X$ we do the matching of Yukawa couplings in the following
way. ${\bf Y_d}$ and ${\bf Y_e}$ are diagonalized.
All the quark mixing included in ${\bf Y_u}$ and all the lepton 
mixing in ${\bf Y_{\nu}}$.
The matrix ${\bf Y_u}$ (symmetrized through a rotation
of the up quark singlets) is then matched to the analogous
matrix above $M_X$. We do {\it not} assume tau-bottom unification. 
${\bf Y_{d/e}}$ is matched to  ${\bf Y_e}$,
what is justified \cite{Illana:2003pj} for both the small and the large values 
of $\tan\beta$ considered.
The matching of SUSY-breaking of the slepton doublet scalar masses is 
${\bf m^2_{\tilde L}}={\bf m^2_{\tilde \Phi}}$ in the studied scenarios
and charged-slepton singlets ${\bf m^2_{\tilde e_R}}={\bf m^2_{\tilde \Psi}}$.

The running introduces three main effects on the flavor structure
of the model, two of them add to the corrections described for
universal masses at $M_X$, whereas the third one is new.
At $M_X$ there appear new off-diagonal terms
in the slepton-doublet mass matrix:
\bea
m^2_{\tilde \Phi\;ij}\approx -{3\over 8\pi^2}
m_0^2 ({\bf Y^{\dagger}_\nu Y_{\nu}})_{ij}\log{M_P \over M_X}\ .
\eea
The second effect is a large mass separation of the third slepton
family produced by tau corrections:
\bea
m^2_{\tilde \Phi\;33}-m^2_{\tilde \Phi\;ii}\approx -{3\over 2\pi^2}
m_0^2\  Y_\tau^2\ \log{M_P \over M_X}\ .
\eea
This mass splitting will introduce off-diagonal mass terms once the
charged-lepton Yukawa matrix is rediagonalized at
low energies.
The final (new) effect has to do with the slepton-singlet mass
matrix. In the minimal SU(5) model this matrix coincides with the
one for up squarks, so it will be affected by large top quark
radiative corrections. At $M_X$ there will be
off-diagonal terms of order
\bea
m^2_{\tilde \Psi\;ij}\approx -{9\over 8\pi^2}
m_0^2 ({\bf Y^{\dagger}_u Y_u})_{ij}\log{M_P \over M_X}\ ,
\eea
where ${\bf Y_u}$ contains the whole CKM rotation (we take
${\bf Y_{d/e}}$ diagonal at $M_X$). This effect was first considered
in \cite{Barbieri:1994pv} and produces off-diagonal terms 
($\delta^{ij}_{RR}$) in 
${\bf m^2_{\tilde e_R}}$, giving the dominant contribution to LFV for
all scenarios except for the first two generations in cases ({\it a1}) and
({\it b1}) which are dominated by the term $m^2_{\tilde L_{12}}$.

To illustrate the relevance of the corrections between $M_P$ and
$M_X$ we give
${\bf m^2_{\tilde L}}$ at $M_Z$ and compare it
with the matrices obtained for universal scalar masses
at $M_X$. We take values of $m_0$ and $m_{1/2}$ 
at $M_P$, which give at $M_X$ similar values to
the ones used in Eqs.~(\ref{eq10},\ref{eq11}):
\bea
&&[m_0=300\mbox{ GeV},\ m_{1/2}=275\mbox{ GeV at }M_P]\nonumber\\
\mbox{({\it a1})}\quad{\bf m^2_{\tilde L}}&=&(353 {\rm\; GeV})^2
\pmatrix{1&{-4\times10^{-4}}&{-7\times10^{-5}}\cr
\quad*\quad&0.997&{-2\times10^{-3}}\cr
*&*&{0.567}\cr} \\
\mbox{({\it b1})}\quad{\bf m^2_{\tilde L}}&=&(349 {\rm\; GeV})^2
\pmatrix{1&{-2\times10^{-4}}&{2\times10^{-4}}\cr
\quad*\quad&0.990&{-10^{-2}}\cr
*&*&{0.989}\cr}
\eea

\begin{figure}
\begin{center}
\begin{tabular}{cc}
\epsfig{file=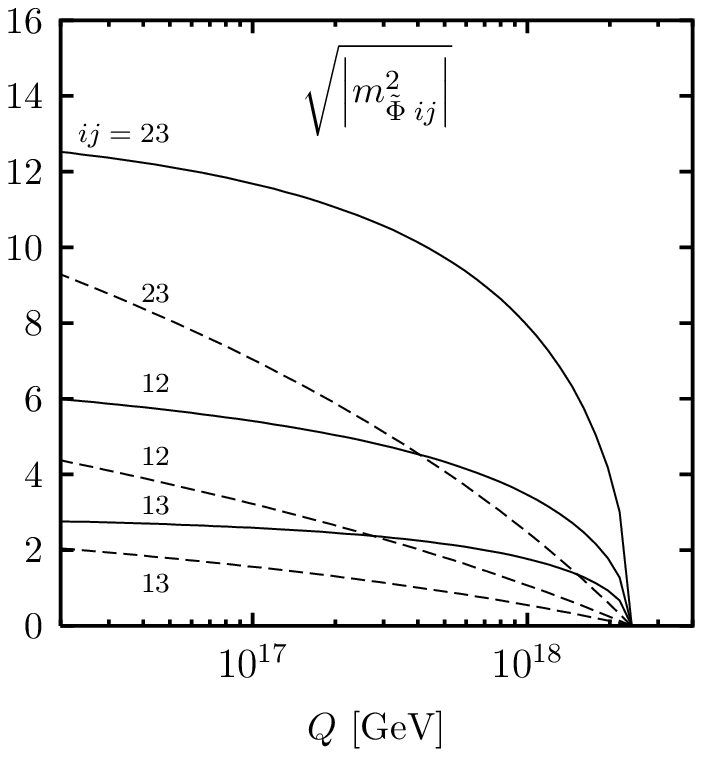,height=6cm} &
\epsfig{file=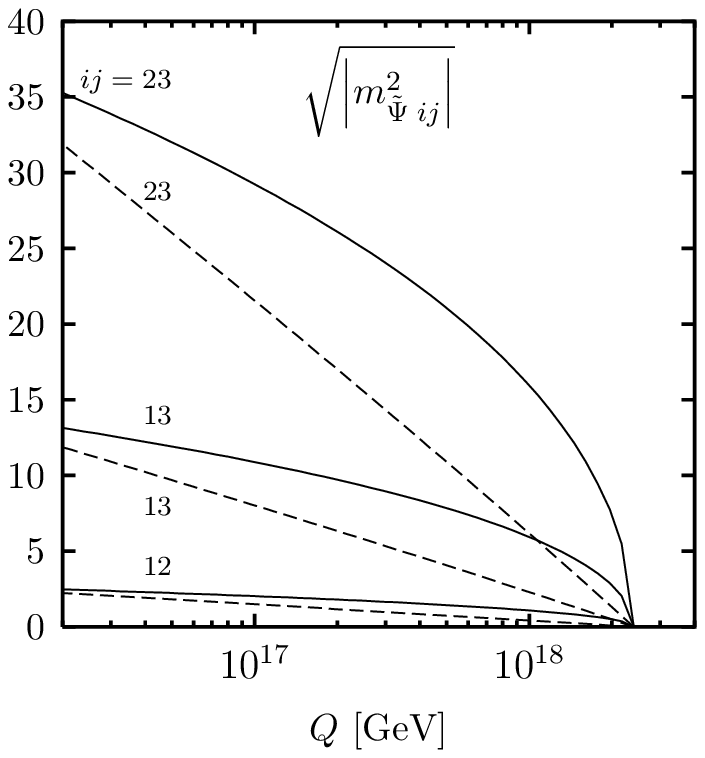,height=6cm}
\end{tabular}
\end{center}
\caption{Off-diagonal terms [GeV] generated in the running of 
${\bf m^2_{\tilde\Phi}}$ and ${\bf m^2_{\tilde\Psi}}$ from $M_P$ to $M_X$. 
Solid lines correspond to no-scale models ($m_0=0$ at $M_P$).
\label{fig1}}
\end{figure}
                                                                                
Finally, contrary to the usual claim, we find that it is not justified
to take universal SUSY-breaking masses at the GUT scale even in the
no-scale models (scalar masses taken zero at Planck scale and radiatively
generated), since off-diagonal terms in the scalar sector induced by RGE 
between $M_P$ and $M_X$ are only reduced only by an 80\% compared to the 
ordinary case (Fig.~\ref{fig1}).

\section{Predictions for LFV decays}

Let us show next how the misalignment between lepton and
slepton families translates into the rate of flavour-violating decays. 
We concentrate on the (best constrained) rare lepton decays. The
present (future) experimental limits are \cite{limits}
\bea
{\rm BR}(\mu\to e\gamma)   & < &
1.2\times 10^{-11} \quad (10^{-14}) \ ,
\\
{\rm BR}(\tau\to e\gamma)  & < &
2.7\times 10^{-6} \ ,
\\
{\rm BR}(\tau\to\mu\gamma) & < &
6\times 10^{-7}   \quad\quad\ (10^{-9}) \ .
\eea
                                                                               
A detailed description of all the diagrammatics involved in these
processes can be found in \cite{Illana:2002tg}.

\begin{figure}
\centerline{\epsfig{file=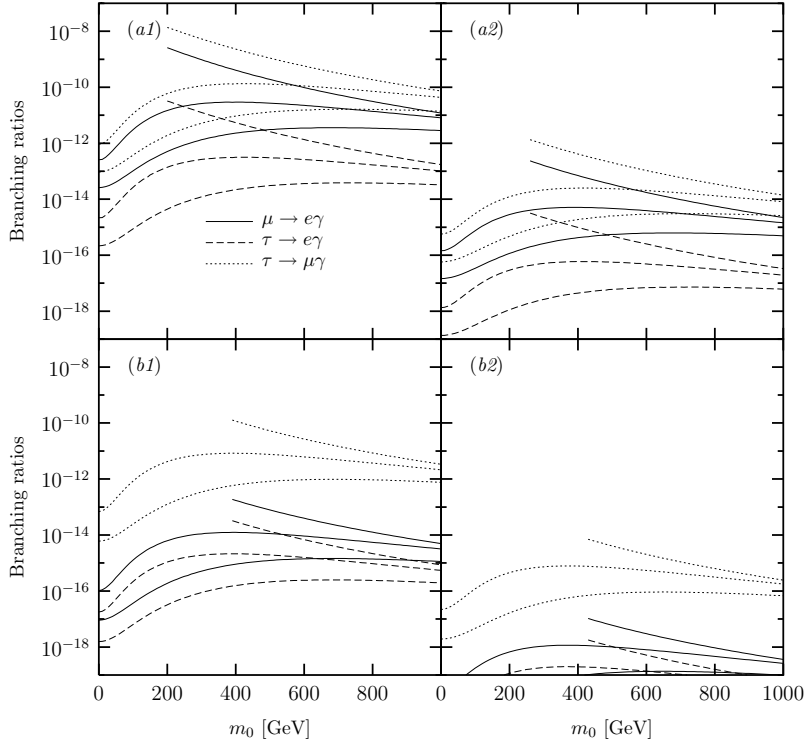,width=0.9\linewidth}}
\caption{Branching ratios of $\ell\rightarrow\ell'\gamma$
for $m_{1/2}=100,300,500$ GeV and different values of the
scalar mass parameter $m_0$ at $M_X$. Cases ({\it a})
and ({\it b}) correspond to $\tan\beta=50,3$, whereas
cases ({\it 1}) and ({\it 2}) correspond to
$M=10^{14},5\times 10^{11}$ GeV, respectively.
Lower values of $m_0$ for $m_{1/2}=100$~GeV give slepton masses
excluded by present bounds.
\label{fig2}}
\end{figure}

The results for the four cases  ({\it a1, a2, b1, b2}) assuming universality
at $M_X$ and choosing $\mu>0$ are summarized in Fig.~\ref{fig2}. 
For each case we plot the branching
ratios of $\mu\to e\gamma$ (solid), $\tau\to e\gamma$ (dashes) and
$\tau\to \mu\gamma$ (dots). In all the cases
the three branching ratios are dominated by diagrams with
exchange of charginos and sneutrinos \cite{Illana:2002tg}.
The angle $\theta_{13}$ is taken zero at $M_X$ for scenarios with low
$\tan\beta$. The rates of $\mu\to e\gamma$ and $\tau\to e\gamma$ in those 
cases would be scaled by a factor $\approx (\theta_{13}/10^{-2})^2$
for values of this angle close to the experimental bound
$\theta_{13}\approx 0.2$. Only the process $\mu\to e\gamma$ and only for
the scenario ({\it a1}) is constrained by present bounds and would be discovered
or ruled out in the near future.

\begin{figure}
\centerline{\epsfig{file=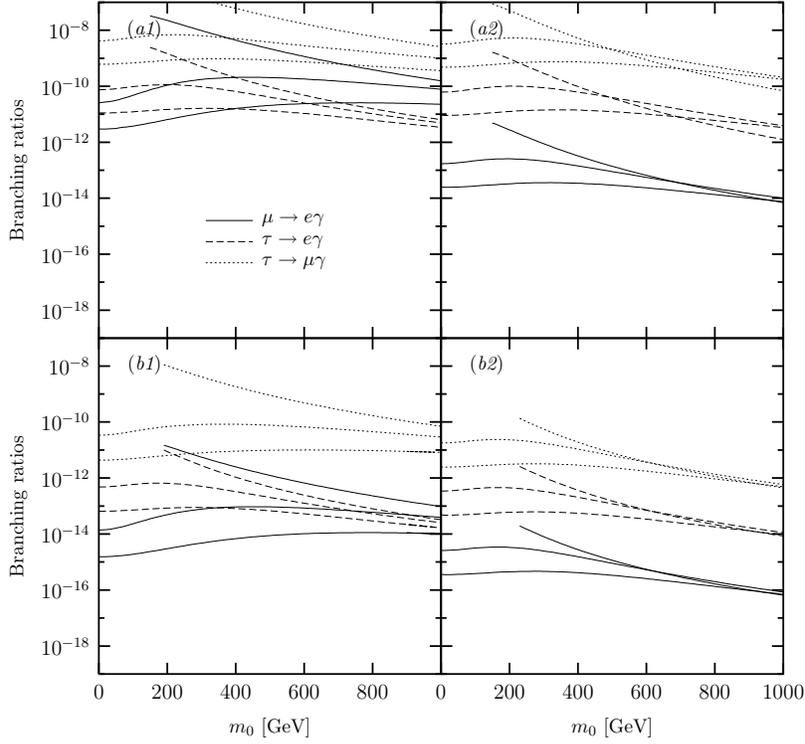,width=0.9\linewidth}}
\caption{The same as in Fig.~\ref{fig2} but assuming universality at $M_P$.
\label{fig3}}
\end{figure}

The corresponding results assuming universality at $M_P$ are shown in
Fig.~\ref{fig3}. It is remarkable the big enhancement of all processes
in scenarios ({\it a2}) and ({\it b2}) and of $\tau\to e\gamma$ and 
$\tau\to\mu\gamma$ in ({\it a1}) and ({\it b1}), due to the dominant 
contribution of the $\delta_{RR}$ insertions radiatively generated by 
top-quark corrections. The rate of $\mu\to e\gamma$ is controlled by 
$\delta^{12}_{LL}$ for ({\it a1}) and ({\it b1}). The latter is obtained
with $\theta_{13}=0$ at $M_X$ and would be scaled, as before, by a factor 
$\approx (\theta_{13}/10^{-2})^2$ for larger values of this angle. 
Comparing Figs.~\ref{fig2} and \ref{fig3} we find that corrections 
from the Planck to the GUT scale can increase BR$(\tau\to \mu\gamma)$ in 
four orders of magnitude (case {\it a2}) or BR$(\mu\to e\gamma)$ in two orders 
of magnitude (case {\it b2}).

To conclude, in models with a gravity-mediated origin of the
SUSY-breaking parameters the lepton flavor problem may be
avoided at the tree level. However, radiative corrections from tau
and neutrino couplings introduce flavor violation at rates
that should be observable in near future experiments.

\vskip 3mm
It is a pleasure to thank the organizers of the Conference
for the enjoyable and stimulating atmosphere of the event
and the kind hospitality in Ustro\'n.
This work has been supported by the Spanish CICYT,
the Junta de Andaluc{\'\i}a and the European Union
under contracts FPA2000-1558, FQM 101,
and HPRN-CT-2000-00149, respectively.

\end{document}